\documentstyle[preprint,prb,aps]{revtex}
\begin{document}
\draft
\title{Far-infrared absorption in parallel quantum wires with weak tunneling}
\author{T. V. Shahbazyan and S. E. Ulloa}
\address{Department of Physics and Astronomy
and Condensed Matter and Surface Science Program, Ohio University,
Athens, OH 45701-2979}
\maketitle

\begin{abstract}
We study collective and single-particle intersubband excitations in a 
system of quantum wires coupled via weak tunneling. For an isolated
wire with parabolic confinement, the Kohn's theorem guarantees that
the absorption spectrum represents a single sharp peak centered at the 
frequency given by the bare confining potential. We show that the
effect of weak  tunneling between two parabolic quantum wires is
twofold: (i) additional peaks corresponding to 
single-particle excitations appear in the absorption spectrum, and
(ii) the main absorption peak acquires a depolarization shift.
We also show that the interplay between tunneling and weak
perpendicular magnetic field drastically enhances the dispersion of
single-particle excitations. The latter leads to a strong damping of
the intersubband plasmon for magnetic fields exceeding a critical
value. 
\end{abstract}
\narrowtext

One-dimensional electronic systems (quantum wires) have
attracted much interest in recent years. Using modern lithographic
techniques, quantum wires are usually fabricated on two-dimensional
electron gas confined to the $z=0$ plane, by
imposing an additional confinement along the $x$ direction. 
Much insight into the electronic properties of such systems is gained
from optical experiments such as far-infrared (FIR) absorption
spectroscopy\cite{han87,als88,bri88,dem88,als89,dem91,dre92}
and inelastic light scattering.\cite{ege90,gon91,gon93}
Owing to the size-quantization of energy levels, the excitation 
spectrum is split into intrasubband and intersubband excitations. 
Extensive theoretical studies of collective excitations (plasmons)
were done both on isolated wires and multiwire 
superlattices.\cite{lee83,que88,li89,que91,li91a,efe91,li91b,hau91} 
In many experiments the confining potential is of parabolic form
to a very good approximation. In this case the generalized Kohn 
theorem\cite{bre89,mak90} guarantees that the frequency of the
absorbed light is precisely equal to the frequency $\omega_0$ of the bare
confining potential and, thus, is independent of interactions and
electron concentration. The resonant frequency $\omega_0$ is, in fact, 
the frequency of the fundamental mode of the intersubband plasmon,
corresponding to a collective motion of the center of mass of the
many-electron system. Since the incident light is coupled to the
center of mass only, the absorption spectrum represents a single sharp
peak centered at this frequency. On the other hand, the frequencies
of single-particle intersubband excitations are renormalized by 
interactions and therefore {\em do} depend on 
concentration.\cite{lau88} These frequencies generally lie below
$\omega_0$; however, they are not visible in FIR absorption experiments.

The nonparabolicity of the confining potential can strongly affect the
absorption spectrum. In particular, a single absorption peak is
split into several peaks corresponding to the frequencies of higher
intersubband resonances.\cite{dre92} Also, the position of the main
absorption peak is shifted from $\omega_0$ by an amount
depending on electron concentration. This depolarization shift,
which is caused by resonant screening, measures, in
fact, the strength of electron-electron interactions in the quantum
wire and can be quite significant if the concentration is
high or the nonparabolicity is strong.\cite{dem88} 
Detailed numerical studies of the effect of nonparabolicity 
on both intra- and intersubband excitations were
performed in Ref. \onlinecite{wen95}.

Usually, absorption experiments are done on quantum wire arrays and
superlatticies rather than on a single wire. Correspondingly, the
properties of collective excitations become dependent also on the
interwire coupling. The latter was shown to be particularly important
for intrasubband plasmons in lateral multiwire 
superlattices.\cite{que91,li91a} In a recent FIR absorption
experiment\cite{dre92} a splitting of the fundamental intersubband
plasmon mode with magnetic field at high electron concentration was
observed. On the other hand, if wires are close enough, so that interwire 
coupling is significant, then the confining potential in each wire is,
strictly speaking, no longer parabolic. It is not {\em a priori}
clear which features of the observed absorption spectrum should be
attributed to the interwire interactions or tunneling, and which are
caused by the nonparabolicity of the confining potential itself.

In this paper we study the effect of interwire tunneling on the
absorption spectrum. The geometry under consideration represents a
pair of identical parallel wires separated by a distance $d$. 
The form of the confining potential for the double-wire system is 
depicted schematically in Fig.~1. It represents a double-well with a
barrier allowing weak tunneling between the wells.
We assume that the two wires are not too close to each other, so that
the confining potential of each wire can still be considered as
parabolic.  We show that the tunneling between two such parabolic
wires brings about the depolarization shift of the intersubband
plasmon frequency.  At the same time, additional peaks appear in the
absorption spectrum, which correspond to the intersubband
single-particle (SP) excitations. Perhaps as the most interesting
effect, we also demonstrate that as a result of level splitting,
caused by tunneling, the absorption spectrum becomes extremely
sensitive to a weak perpendicular magnetic field. In particular, for
fields larger than a critical value, the plasma mode is shown to be
strongly damped by single-particle excitations, providing a unique
experimental signature of the effect. We also argue that a similar,
but weaker, effect should be present in a system of coupled quantum
wells. 

To start with, the bare confining potential for each wire is parabolic,  
$m\omega_0^2(x\pm d/2)^2/2$ ($m$ is electron mass), with the upper
(lower) sign corresponding to the left (right) wire. The energy
spectrum of free electrons in the  $i$th wire ($i=1,2$ for the left
and right wire, respectively) is given by $E_{in}(k)=E_{in}+p^2/2m$, 
where $p$ is the momentum along the wire and $E_{in}$ is the
energy of the $n$th subband (we set $\hbar=1$). In the following we
assume that concentration is low and take into account only two (with
$n=0,1$) subbands in each wire with only the lowest subband occupied in
the ground state. The wave-functions in each wire are represented by
the products 

\begin{eqnarray}\label{psi}
\psi_{pn}^i({\bf r})={e^{ipy}\over \sqrt{L_y}}\varphi_n^i(x),
\end{eqnarray}
where $L_y$ is the normalization length. The tranverse wave-functions
$\varphi_n^i(x)$ are harmonic 
oscillator wave-functions centered at the potential minima:
$\varphi_n^{i}(x)=\varphi_n(x\pm d/2)$ with upper (lower) sign
corresponding to $i=1~(2)$. The free tight-binding Hamiltonian for the
double-wire system with tunneling has the form

\begin{eqnarray}\label{h0}
H_0=\sum_{in}E_{in}(p)a^{\dagger}_{in}(p)a_{in}(p) 
+\sum_{n}t_n(p)[a^{\dagger}_{1n}(p)a_{2n}(p)+h.c.],
\end{eqnarray}
were $a^{\dagger}_{in}(p)$ is creation operator and $t_n(p)$
is the tunneling amplitude. We assume that the tunneling is weak:
specifically, the amplitude $t_n(p)$ is much smaller than the subband
separation, $\Delta$, for all $|p|\leq p_F$ ($p_F$ is the Fermi
momentum). We also assume that the energy separation between the Fermi
level and the bottom of the $n=1$ subbands is also of the order of
$\Delta$. In this case we can ignore the tunneling between $n=0$
subbands and set $t_0(p)=0$ in the rest of the paper. 
We consider identical wires in what follows so that 
$E_{1n}(p)=E_{2n}(p)=E_{n}(p)$.

The dependence $t_1(p)$ reflects the fact that the tunneling amplitude
depends on energy and, thus, is an increasing function of $p$. The
specific form of this dependence is determined by the shape of the 
confining potential in the tunneling region. For weak tunneling, 
the amplitude $t_1(p)$ can be calculated quasiclassically\cite{}

\begin{eqnarray}\label{t}
t_1(p)={\hbar\omega_0\over 2\pi}
\exp\Biggl[-{1\over\hbar}\int_{-s}^{s}\sqrt{V(x)-E_1(p)}dx\Biggr],
\end{eqnarray}
where $s(p)$ is the classical turning point to be found from equation
$V(s)-E_1(p)=0$. In order to have an analytical expression for
$t_1(p)$, we adopt the following approximation for the potential in
the tunneling region 

\begin{eqnarray}\label{vtun}
V(x)=V_0-m\Omega^2x^2/2,
\end{eqnarray}
where $\Omega$ is the curvature of the potential and $V_0$
is its height. With such $V(x)$ we obtain from (\ref{t})

\begin{eqnarray}\label{tp}
t_1(p)={\hbar\omega_0\over 2\pi}
\exp\Biggl[-{V_0-E_1(p)\over\gamma}\Biggr], 
\end{eqnarray}
where $\gamma=\Omega/\pi$. This expression can be more conveniently
presented as $t_1(p)=te^{p^2/p_0^2}$, where $t$ is tunneling
amplitude at the bottom of the $n=1$ subband and the parameter
$p_0=\sqrt{2m\Omega}/\pi$ characterizes the increase in tunneling 
with increasing $p$. Now, the $n=1$ subbands in the wires form ``new''
subbands due to the tunneling, with energies  

\begin{eqnarray}\label{epm}
E_{\pm}(p)=E_{1}+p^2/2m \pm t_1(p).
\end{eqnarray}
As can be seen, the subband splitting, $2t_1(p)=2te^{p^2/p_0^2}$,
increases with $p$.  Correspondingly, the separation between lower
(assumed unchanged here) and upper (new) subbands become $p$-dependent: 
$E_{\pm}(p)-E_{0}(p)=\Delta \pm t(p)$. This means that frequencies of
vertical  (with no momentum transfer) intersubband SP transitions are
now distributed in a finite interval. 

The total Hamiltonian of the system with both intra- and interwire
electron-electron interactions includes also the term

\begin{eqnarray}\label{hint}
H_{int}={1\over 2}\sum v_{klmn}^{ij}(p)
a^{\dagger}_{ik}(p)a_{il}(p)a^{\dagger}_{jm}(p-q)a_{jn}(p+q).
\end{eqnarray}
Here $v_{klmn}^{ij}(p)$ are the matrix elements of the Coulomb potential
calculated from transverse wave-functions (\ref{psi})

\begin{eqnarray}\label{vmat}
v_{klmn}^{ij}(p)=\int dx dx'
\varphi_{k}^i(x)\varphi_{l}^i(x)
v(x-x',p)
\varphi_{m}^j(x')\varphi_{n}^j(x'),
\end{eqnarray}
where $v(x,p)=(2e^2/\kappa)K_0(px)$ is the Fourier transform of
$v(x,y)=e^2/\kappa \sqrt{x^2+y^2}$ with 
respect to $y$ ($K_0$ is the modified Bessel function and $\kappa$ is
the dielectric constant). 

The absorption power of incident light polarized across the wires is
related to the conductivity $\sigma(\omega)$ by

\begin{eqnarray}\label{power}
{\cal P}(\omega)={1\over 2}\mbox{Re}[{\cal E}^2\sigma(\omega)],
\end{eqnarray}
where ${\cal E}$ is external electric field. In the following we
evaluate $\sigma(\omega)$ within the random-phase approximation (RPA). 
The system described by the Hamiltonian $H=H_0+H_{int}$ contains both
intra- and intersubband excitations. It can be shown, however, that
only the latter contribute to the response to a spatially homogeneous
electric field.\cite{que88} In this case the expression for the 
conductivity reads 

\begin{eqnarray}\label{sig}
\sigma(\omega)=e^2\omega
\mbox{Im}\sum_{ij}x_{01}^i \Pi^{ij}(\omega) x_{01}^j
\end{eqnarray}
where $x_{01}^i$ is the intersubband matrix element of $x$ in the
$i$th wire. The matrix {\boldmath{$\Pi$}} is related to matrix elements of 
interaction (\ref{vmat}) as (in matrix form)

\begin{eqnarray}\label{pi}
\mbox{\boldmath{$\Pi$}}(\omega)=
\mbox{\boldmath{$\chi$}}(\omega) 
[1-\mbox{\boldmath{$v$}}_{1010}(0)
\mbox{\boldmath{$\chi$}}(\omega)]^{-1},
\end{eqnarray}
where $\mbox{\boldmath{$\chi$}}(\omega)$ is the intersubband
polarization of free electrons which, in general, is also a matrix in
the wire indexes. It can be shown further that in the model with the tunneling
between $n=1$ subbands only, the polarization matrix is diagonal, with equal
elements for identical wires: 
$\chi^{ij}(\omega)=\delta_{ij}\chi(\omega)$. The function $\chi(\omega)$
can be easily calculated using standard Matsubara
techniques.\cite{mahan} The result is 

\begin{eqnarray}\label{chi}
\chi(\omega)=2\int {dp\over 2\pi}n_0(p)
{\omega-\Delta\over (\omega-\Delta)^2-t_1^2(p)}+
(\omega\leftrightarrow -\omega),
\end{eqnarray}
where $n_0(p)$ is the Fermi function (we assume concentration in both
wires equal and measure the Fermi level from the bottom of the $n=0$
subband). The poles of $\Pi_{ij}$, given by secular equation 
$\mbox{det}[1-v_{1010}(0)\chi(\omega)]=0$, determine the frequencies
of {\em in}- and {\em out}-of-phase intersubband plasmon modes. For
identical wires, the two modes get decoupled and the latter equation
takes the form $(1-v^{+}\chi)(1-v^{-}\chi)=0$, where

\begin{eqnarray}\label{vpm} 
v^{\pm}=v_{1010}^{11}(0)\pm v_{1010}^{12}(0).
\end{eqnarray} 
Note that in this case only the {\em in}-phase mode, described by
equation  

\begin{eqnarray}\label{plas}
1-v^{+}\chi(\omega)=0,
\end{eqnarray}
contributes to the absorption. Indeed, using $x_{01}^i=l/\sqrt{2}$ with
$l=1/\sqrt{m\omega_0}$, one can easily obtain from (\ref{sig}) and
(\ref{pi}) after simple algebra that

\begin{eqnarray}\label{sigma}
\sigma(\omega)=e^2l^2\omega\mbox{Im}
{\chi(\omega)\over 1-v^{+}\chi(\omega)}=e^2l^2\omega
{\mbox{Im}\chi(\omega)\over [1-v^{+}\mbox{Re}\chi(\omega)]^2
+[v^{+}\mbox{Im}\chi(\omega)]^2}.
\end{eqnarray}

Generally, in order to make RPA consistent with Kohn's theorem, it is
necessary to include the effects of screening by adding 
the self-consistent potential due to the electron-electron 
interaction to the bare confining potential.\cite{bre90} This leads, in
particular, to the lowering of the the subband separation and, hence, 
to a shift of the SP excitation frequencies from
$\Delta=\omega_0$ to 
\begin{eqnarray}\label{del}
\Delta=\omega_0 -U,
\end{eqnarray}
where $U$ is the Hartree energy. This change in $\Delta$ cancels, in turn, 
the depolarization shift of the intersubband plasmon frequency, which
otherwise is present in the ``naive'' RPA for the parabolic
confinement. For low concentrations, with only the lowest subband
occupied, the screening is weak\cite{lau88} and can be treated
perturbatively.\cite{hau91} In this case one can use the unperturbed
wave-functions (\ref{psi}) in calculating the matrix elements
(\ref{vmat}) and substitute for subband separation 
$\Delta=\omega_0+\Sigma_1-\Sigma_0$, where $\Sigma_n$ is the first-order
Hartree self-energy for the $n$th subband. Having in mind that 
the $\Sigma_n$ in each wire includes contributions from
both wires, the corresponding self-energies are easily found to be

\begin{eqnarray}\label{self}
\Sigma_1=[v_{1100}^{11}(0)+v_{1100}^{12}(0)]N, ~~~~
\Sigma_0=[v_{0000}^{11}(0)+v_{0000}^{12}(0)]N, 
\end{eqnarray}
where $N=2p_F/\pi$ is the 1D electron concentration. Note that although
each $\Sigma_n$ diverges, the difference $U=\Sigma_0-\Sigma_1$ is
finite. Using the identity

\begin{eqnarray}\label{ident}
v_{0000}^{ij}(p)-v_{1100}^{ij}(p)=v_{0101}^{ij}(p),
\end{eqnarray}
which can be proven using the properties of the harmonic oscillator
wave-functions in (\ref{vmat}), one then finds 

\begin{eqnarray}\label{U}
U=[v_{1010}^{11}(0)+v_{1010}^{12}(0)]N=v^{+}N,
\end{eqnarray}
where the analytical form of $v_{1010}^{ij}(0)$ follows from
(\ref{vmat})

\begin{eqnarray}\label{vval}
v_{1010}^{11}(0)={e^2\over \kappa},~~~~
v_{1010}^{12}(0)={e^2\over \kappa}\int_0^{\infty}dxxe^{-x^2/2}\cos(xd/l).
\end{eqnarray}
For the interwire distance larger than the wire thickness, $d\gg l$,
the integral in (\ref{vval}) can be estimated as $l^2/d^2$. In this
case we have $U\simeq (e^2N/\kappa)(1+l^2/d^2)\simeq e^2N/\kappa$,
i.e., the Hartree energy is dominated by the intrawire interaction.

It can be easily checked that in the absence of tunneling, the
solution $\omega_p$ of plasmon equation (\ref{plas}) with $\Delta$ 
given by (\ref{del}) and (\ref{U}) is indeed $\omega_p=\omega_0$. Since
$\mbox{Im}\chi(\omega)=0$ for all $\omega$ except 
$\omega=\Delta\neq\omega_0$, the conductivity (\ref{sig}) represents a
single sharp peak centered at $\omega_0$.

Let us now first consider the effect of tunneling on SP excitations. With
$t_1(p)$ given by (\ref{tp}), $\mbox{Im}\chi(\omega)$ can be
explicitly evaluated from (\ref{chi}) with the result

\begin{eqnarray}\label{imchi}
\mbox{Im}\chi(\omega)={\pi\over 4}{p_0\over p_F}
{N\over |\omega-\Delta|}\ln^{-1/2}
\Biggl|{\omega-\Delta \over t}\Biggr|,~~~~
\mbox{for}~~~t<|\omega-\Delta|< te^{p_F^2/p_0^2},
\end{eqnarray}
and $\mbox{Im}\chi(\omega)=0$ otherwise. Eq.~(\ref{imchi}) shows that
the tunneling leads to a non-zero $\mbox{Im}\chi(\omega)$ in two 
frequency intervals. This is related to the fact, already mentioned
above, that the dependence of level splitting (\ref{epm}) on momentum
along the wires leads to a continuum of SP excitation frequencies even
at zero momentum transfer. The width of each interval is determined by
the difference between level splitting at the bottom ($p=0$) and at
the edge ($p=p_F$) of the $n=1$ subbands. Note that the divergency of
$\mbox{Im}\chi(\omega)$ at $\omega=\Delta\pm t$ is related to the
divergency of the 1D density of states at $p=0$. Since
$\mbox{Im}\chi(\omega)$ itself is finite in the interval (\ref{imchi}), 
the conductivity (\ref{sig}) should exhibit two peaks corresponding to
resonant absorption of incident light at the allowed SP excitations
frequencies.

Let us now turn to the main absorption peak corresponding to the
intersubband plasmon. Unlike the interwire interaction, the tunneling
{\em does} shift the frequency of the {\em in}-phase mode. (Note that
the frequency of the {\em out}-of-phase mode is shifted from
$\omega_0$ by the amount $-2v_{1010}^{12}(0)N$ even in the absence of
tunneling). One can estimate this shift for the experimentally
relevant case of the splitting being less than the Hartree energy. 
For $U\gg t(p_F)$, we expect that the depolarization shift 
$\delta\omega\equiv\omega_p-\omega_0\ll U$. Then for the relevant
$\omega$ the integrand in the expression (\ref{chi}) for $\chi$  can be
expanded and up to first order in tunneling one gets

\begin{eqnarray}\label{chiexp}
\chi(\omega)={N\over \omega-\Delta}+
{N\overline{t^2}\over (\omega-\Delta)^3},
\end{eqnarray}
where
\begin{eqnarray}\label{tms}
\overline{t^2}={1\over p_F}\int_0^{p_F}dpt^2(p),
\end{eqnarray}
is the mean square tunneling. Substituting (\ref{chiexp}) into 
(\ref{plas}) and solving for $\omega$ we then obtain the following
estimate for the depolarization shift

\begin{eqnarray}\label{dep}
\delta\omega
\simeq{\overline{t^2}\over U}.
\end{eqnarray}
We see that the depolarization shift changes with tunneling quadratically. 
The condition $U\gg t(p_F)$, under which Eq.~(\ref{dep}) was derived,
also guarantees that the allowed SP excitation frequencies lie below
$\omega_p$, so that $\mbox{Im}\chi(\omega_p)=0$. In other words, the
absorption spectrum represents a sharp peak at $\omega=\omega_p$ well
separated from two smaller and wider peaks corresponding to the SP
excitations.

In Fig.~2 we present results of numerical calculations for the
depolarization shift of the intersubband plasmon frequency.
We take $\omega_0=5.2$ meV in accordance to Ref.~\onlinecite{gon91},
where parabolic wires with only the lowest subband occupied were
fabricated. We plot $\delta\omega$ versus tunneling amplitude at the
bottom of the subband, $t$, for three different values of concentration,
which correspond to a Hartree energy $U\simeq 0.3\omega_0$ (with
$\kappa=12$ corresponding to GaAs). The parameter 
$\Omega=\pi^2p_0^2/2m$, characterizing the difference in splitting
between the bottom and the edge of the subbands, is taken as
$\Omega=0.25\omega_0$. This difference is then of the order of
tunneling itself, so that $\overline{t^2}\sim t^2$. It can be seen
that the dependence of the depolarization shift on $t$ is quite close
to quadratic.

In Fig.~3 the conductivity is plotted in units of
$\sigma_0=e^2N/m\omega_0$ for two different values of $t$ and the same
$\omega_0$ and $\Omega$ as above. The two smaller left peaks
correspond to the SP excitations, while the sharp peak at the right
corresponds to the intersubband {\em in}-phase plasmon (we have
introduced a small phenomenological $\mbox{Im}\chi$ in order to make the
latter peak visible on the graph). It can be seen that the SP peaks
are asymmetric and become larger and higher with increasing $t$. With
the parameters used, the width of the peaks, $t(p_F)$, as well as their
separation, $2t$, is less than $U$ so that the SP and plasmon peaks are
well separated. Note, however, that since since the SP futures are
much smaller than the plasmon peak, they would not be seen in
experiments where inhomogeneous broadening may be of the same order
as the SP peaks.

Let us now study the effect of a perpendicular magnetic field on the
absorption spectrum in the presence of tunneling. We assume that the
magnetic field is weak so that $\omega_c\ll\omega_0$, where
$\omega_c=eB/mc$ is the cyclotron frequency. Choosing the vector-potential
in the Landau gauge, ${\bf A}=(0,Bx,0)$, the wave-functions again
can be presented as products of the form (\ref{psi}). Note that the
transverse functions in the presence of the magnetic field acquire a 
dependence on $p$: $\varphi_{np}^{i}(x)=\varphi_{n}(x\pm d/2 +\alpha p)$, 
where $\alpha=\omega_c/m(\omega_0^2+\omega_c^2)$. For
$\omega_c\ll\omega_0$ and low concentrations one has 
$\alpha p\simeq (\omega_c/\omega_0)pl^2\ll l$, and the wave-functions
are basically unaffected by the magnetic field.\cite{li91b}
Correspondingly, the matrix elements of the interaction and, hence,
the subband separation $\Delta$ are given by their zero-field
expressions (\ref{U}), (\ref{vval}), and (\ref{del}). On the other
hand, since the minima of confining potential of the double-wire
system are centered at different values of $x$, the magnetic field
affects the $p$-dependence of the free electron energies. With the
above choice of the gauge, the eigenenergies in different wires are
given by

\begin{eqnarray}\label{Eshift}
E_{in}(p)=E_n+{1\over 2m}\Biggl(p\pm {d\over 2l_c^2}\Biggr)^2,
\end{eqnarray}
with upper (lower) sign corresponding to $i=1$ (2) and
$l_c=1/\sqrt{m\omega_c}$ being the magnetic length. In
Eq.~(\ref{Eshift}) we omitted the factor 
$\omega_0^2/(\omega_0^2+\omega_c^2)\simeq 1$ in front of the second
term. Thus, the only effect of a weak magnetic field is to shift the
momenta of electrons in different wires in {\em opposite}
directions. In the absence of tunneling, this constant shift does not
play any role. However, with the tunneling turned on the situation
becomes completely different. It is important that the confining
potential is $y$-independent so that the tunneling amplitude 
$t(p)$ connects states with the same momentum. Diagonalizing the
free Hamiltonian (\ref{h0}) with $E_{in}(p)$ from (\ref{Eshift}), one
obtains for the new upper subbands

\begin{eqnarray}\label{epmmag}
E_{\pm}(p)=\overline{E}_{1}(p)\pm \lambda(p),
\end{eqnarray}
where $\overline{E}_{1}(p)=[E_{11}(p)+E_{21}(p)]/2$ and $\lambda(p)$
is the field-dependent splitting,

\begin{eqnarray}\label{lam}
\lambda(p)=\sqrt{\xi^2(p)+t_1^2(p)},
\end{eqnarray}
with $\xi(p)=[E_{11}(p)-E_{21}(p)]/2=pd\omega_c/2$. For weak
tunneling, one can still use the expression (\ref{tp}) for $t(p)$ with
$\overline{E}_{1}(p)$ instead of $E_{1}(p)$.

The intersubband polarization $\chi(\omega)$ in the presence of
both magnetic field and tunneling, calculated from the Hamiltonian
(\ref{h0}) and (\ref{Eshift}), now has the form 

\begin{eqnarray}\label{chimag}
\chi(\omega)=2\int {dp\over 2\pi}n_0(p-d/2l_c^2)
{\omega-\Delta-2\xi(p)
\over [\omega-\Delta-\xi(p)]^2-\lambda^2(p)}+
(\omega\leftrightarrow -\omega).
\end{eqnarray}
The argument of the Fermi function in (\ref{chimag}) implies that the
momentum integration is now carried out in the interval 
$-\pi N/2+d/2l_c^2\leq p\leq \pi N/2+d/2l_c^2$. In other words,
magnetic field shifts the Fermi momenta to 

\begin{eqnarray}\label{pfpm}
p_F^{\pm}=\pm p_F+d/2l_c^2.
\end{eqnarray}
Note that the while the expression
(\ref{chimag}) gives $\chi(\omega)$ in the right wire, the
polarizations in both wires are, in fact, equal. Indeed,
$\chi(\omega)$ in the left wire differs from (\ref{chimag}) by the
opposite sign in front of $d$; then it can be brought to the form
(\ref{chimag}) by a change in the integration variable
$p\leftrightarrow -p$. It can be readily seen with the help of
(\ref{lam}) that in the absence of tunneling the polarization
(\ref{chimag}) is independent of magnetic field. 

The combined effect of magnetic field and tunneling on the SP
excitations is illustrated in Fig.~4. At zero tunneling 
[see Fig.~4(a)], the energy levels in each wire represent two
parabolas, corresponding to $n=0$ and $n=1$ subbands, separated by 
$\Delta$. The subbands minima in different wires are shifted in 
$p$-space in opposite directions, according to Eq.~(\ref{Eshift}). The
SP vertical transitions, shown by arrows, can occur within
each wire only and have frequency $\Delta$. With the tunneling turned
on [see Fig.~4(b)], the upper subbands split and form new subbands, 
according to Eq.~(\ref{epmmag}). Since each of the new subbands is a
``mixture'' of the two ``old'' subbands with different minima, 
the range of allowed SP transitions now have much higher
frequencies. The maximal SP excitation frequencies, 
$\omega_{max}=\Delta+\lambda(p_F^{+})+\xi(p_F^{+})$, are due to 
transitions occurring at the subband edge, that is at $p=p_F^{+}$
($p=-p_F^{+}$) for the right (left) wire. Since the separation (in
$p$-space) between two parabolas, corresponding to different wires,
increases with magnetic field, the difference between energies at
the {\em same} $p$ also increases. At magnetic fields such that
$\xi(p_F^{+})\gg t(p_F^{+})$ [but still  $\xi(p_F^{+})\ll \omega_0$]
one has $\lambda(p_F^{+})\simeq \xi(p_F^{+})=p_F^{+}d\omega_c/2$, 
so that the increase in SP excitation frequencies due to the tunneling, 
$\omega_{max}-\Delta\simeq p_F^{+}d\omega_c$, can be much larger than the
tunneling amplitude itself. 
On the other hand, the depolarization shift of the intersubband
plasmon is determined by the tunneling, as we have seen above, so
that $\delta\omega\equiv\omega_p-\omega_0\ll p_F^{+}d\omega_c$. As a
result, with an increasing magnetic field the SP excitation
frequencies can become higher than the plasmon frequency. In other
words, for such fields the intersubband plasmon is 
damped by the SP excitations. The critical field at which the plasmon
frequency enters into the region of SP excitations can be determined
by equating $\omega_{max}$ and $\omega_p$. Having in mind that
$\Delta=\omega_0-U$ and $\delta\omega\ll U$, the condition
$\omega_{max}=\omega_p$ can be written as 

\begin{eqnarray}\label{crit}
p_F^{+}d\omega_c\simeq U.
\end{eqnarray}
Note that $p_F^{+}$ itself depends on magnetic field as well as on
interwire distance.

The damping of the intersubband plasmon by SP excitations leads to a 
strong broadening of the absorption peak as the magnetic field exceeds its
critical value. In order to estimate the width of the peak, one should
evaluate  $\mbox{Im}\chi(\omega)$ for frequencies
$\omega\simeq\omega_p$. Decomposing the integrand in (\ref{chimag})
as 

\begin{eqnarray}\label{decompos}
{\omega-\Delta-2\xi(p)
\over [\omega-\Delta-\xi(p)]^2-\lambda^2(p)}
={1-[\lambda(p)-\xi(p)]/2\lambda(p)
\over \omega-\Delta-\xi(p)+\lambda(p)}+
{\lambda(p)-\xi(p)\over 2\lambda(p)}
{1 \over \omega-\Delta-\xi(p)-\lambda(p)},
\end{eqnarray}
and noticing that only the second term contributes for
$\omega\simeq\omega_p$, the expression for $\mbox{Im}\chi(\omega)$
takes the form

\begin{eqnarray}\label{imchimag}
\mbox{Im}\chi(\omega)=\int_{p_F^{-}}^{p_F^{+}}dp
{\lambda(p)-\xi(p)\over 2\lambda(p)}
\delta[\omega-\Delta-\xi(p)-\lambda(p)].
\end{eqnarray}
For relevant momenta, $\lambda(p)$ can be expanded as 
$\lambda(p)\simeq \xi(p) +t_1^2(p)/2\xi(p)$ so that
$[\lambda(p)-\xi(p)]/2\lambda(p)\simeq t_1^2(p)/4\xi^2(p)$. Then
neglecting the second term of the expansion of $\lambda(p)$ in the
argument of the $\delta$-function and performing the integral we obtain

\begin{eqnarray}\label{imchiest}
\mbox{Im}\chi(\omega_p)\simeq {t_1^2(U/d\omega_c)\over d\omega_c U^2},
\end{eqnarray}
where we replaced $\omega_p-\Delta$ by $U$. The argument of the
tunneling amplitude in (\ref{imchiest}) is smaller than $p_F^{+}$, in
accordance with (\ref{crit}). Substituting (\ref{imchiest}) into
(\ref{sigma}) we obtain the width $\Gamma$ of the conductivity peak

\begin{eqnarray}\label{width}
\Gamma\simeq {t_1^2(U/d\omega_c)\over dN\omega_c}.
\end{eqnarray}
Correspondingly, the height of the peak is reduced, the peak-value of
conductivity now being $\sigma(\omega_p)\simeq e^2N/m\Gamma$.

In Fig.~5 we plot regions of allowed SP excitation frequencies versus
magnetic field, obtained by numerical evaluation of
$\mbox{Im}\chi(\omega)$. With $\omega_0=5.2$ meV,
$\Omega/\omega_0=3.0$, $d/l=2.0$, and $t/\omega_0=0.02$, the SP
excitation frequencies at zero field lie well below the plasmon
frequency $\omega_p\simeq \omega_0$. It can be seen that with
increasing field the upper boundary of SP frequencies also increases
and reaches $\omega_0$ at $\omega_c\simeq 0.17 \omega_0$, which
correspond to a magnetic field $B\simeq 0.6$ T. In Fig.~6 we plot the
conductivity for the same parameters and for magnetic fields close to
the critical value. For $\omega_c= 0.15 \omega_0$, which is just below
the critical field, the plasmon peak is separated from the SP peaks,
only one of which is visible for the parameters chosen. For $\omega_c=
0.2 \omega_0$, just above the critical field, the two peaks merge and
the absorption spectrum represents a single broadened peak. 

In conclusion, we studied the effect of weak tunneling on intersubband
SP and collective excitations in a system of two quantum
wires with parabolic confinement. We have shown that the frequency of
{\em in}-phase intersubband plasmon acquires a depolarization shift
depending quadratically on the tunneling amplitude. The important
distinction between tunneling-induced shift and shift caused by the
nonparabolicity of the confining potential is that they depend
differently on electron concentration. While the latter increases with
the concentration for most types of nonparabolicity,\cite{wen95} the
former is, in fact, inversely proportional to concentration [see
Eqs.~(\ref{dep}) and (\ref{U})]. It should be emphasized, however,
that this is true only for low concentrations, when the Hartree energy
is small and the subband spacing only weakly depends on
density,\cite{lau88} so that the tunneling amplitude $t$ is nearly
density-independent. We have also shown that the SP excitation
frequencies lie in two intervals with the width strongly dependent on 
parameters of the system, and that these excitations
could, in principle, show up in FIR absorption experiments.

The most interesting feature of the system considered is its anomalous
sensitivity to a weak magnetic field. We demonstrated that there is a
critical magnetic field at which intersubband plasmon becomes strongly
damped by SP excitations. Remarkably, this damping occurs at zero
momentum, so that it should reveal itself in a sudden broadening of
the absorption peak as the magnetic field reaches its critical
value. The critical value is determined by concentration and interwire
distance and does not depend on tunneling itself. The tunneling,
however, does determine  the width of the peak. The latter, in fact,
characterizes the ``mixing'' of the ``old'' $n=1$ subbands, which make
up the new ones.  Clearly, with increasing field this mixing should
become weaker so that the absorption peak should become narrower
again. Such a simple dependence holds, however, for weak enough
magnetic fields, satisfying the condition
$(\omega_c/\omega_0)lp_F^{+}\ll 1$. For higher fields the zero-field
approximation for the wave functions is not valid anymore, so that the
subband separation and, hence, the tunneling amplitude $t$ also become
field-dependent. 

Finally, we note that the considerations above can be
straightforwardly generalized to a system of two quantum wells with
parabolic confinement.\cite{har96} In particular, it can be shown that
the depolarization 
shift of the intersubband plasmon frequency is still given by 
Eq.~(\ref{dep}), where the Hartree energy (at low concentration) is
now $U\sim (e^2/\kappa)lN$, $N=p_F^2/2\pi$ being here the 2D electron
concentration. The {\em in-plane} magnetic field shifts 
the centers of the energy paraboloids, corresponding to 
different planes, in opposite directions in the $p$-space
by $\pm d/2l_c^2$. This again leads to the enhancement of
the SP excitations frequencies, resulting in a damping of the plasmon 
at critical field given by Eq.~(\ref{crit}). However, the {\em width}
of the absorption peak is reduced as compared to that in the
double-wire system. In a similar manner, one can estimate the width in
the double-well system as 
\begin{eqnarray}\label{well}
\Gamma\sim {t^2\over dp_F\omega_c}
\sqrt{(p_F^{+}-U/d\omega_c)/p_F},
\end{eqnarray}
with $p_F^{+}$ given by Eq.~(\ref{pfpm}). This expression differs from
Eq.~(\ref{width}) by a factor, which is small when the
magnetic field is close to critical. Physically, the weaker damping 
of plasmon in double-well versus double-wire system originates from
the 2D versus 1D nature of the energy surface in the $p$-space, since
a smaller {\em fraction} of the SP excitations in the former system
have frequencies close to the plasmon frequency and thus participates
in the damping.
 
\acknowledgments

We thank W. Hansen for helpful discussions. This work was supported in
part by US Department of Energy Grant No. DE-FG02-91ER45334 and the
A. v. Humboldt Foundation (SEU).

\begin{figure}
\caption{Schematic picture of the confining potential for the
double-wire system. Each well has two energy levels  
separated by $\Delta=E_1-E_0$.}
\end{figure}

\begin{figure}
\caption{Calculated depolarization shift of the intersubband
plasmon at \protect$\omega_0=5.2$ meV is shown 
for $N=1.0\times 10^5\mbox{ cm}^{-1}$ (dashed line),
$N=1.25\times 10^5\mbox{ cm}^{-1}$ (solid line), and
$N=1.5\times 10^5\mbox{ cm}^{-1}$ (long-dashed line).}
\end{figure}

\begin{figure}
\caption{Calculated conductivity at 
\protect$N=1.0\times 10^5\mbox{ cm}^{-1}$
is shown for (a) $t=0.05\omega_0$ and (b) $t=0.1\omega_0$.}
\end{figure}

\begin{figure}
\caption{Schematic picture of subbands in the magnetic
field in the absence (a) and  presence (b) of tunneling. Arrows
indicate allowed SP transitions for the right wire.}
\end{figure}

\begin{figure}
\caption{Magnetic-field dispersion of SP excitations for
\protect$N=1.0\times 10^5\mbox{ cm}^{-1}$ and $t=0.02\omega_0$. The
shaded areas indicate ranges of frequencies given by the condition
\protect$\mbox{Im}\chi(\omega)\neq 0$. Critical field corresponds to
$\omega_c/\omega_0\simeq 0.17$.}
\end{figure}

\begin{figure}
\caption{Calculated conductivity at 
\protect$N=1.0\times 10^5\mbox{ cm}^{-1}$ and $t=0.02\omega_0$
is shown for (a) $\omega_c=0.15\omega_0$ and 
(b) $\omega_c=0.2\omega_0$.}
\end{figure}

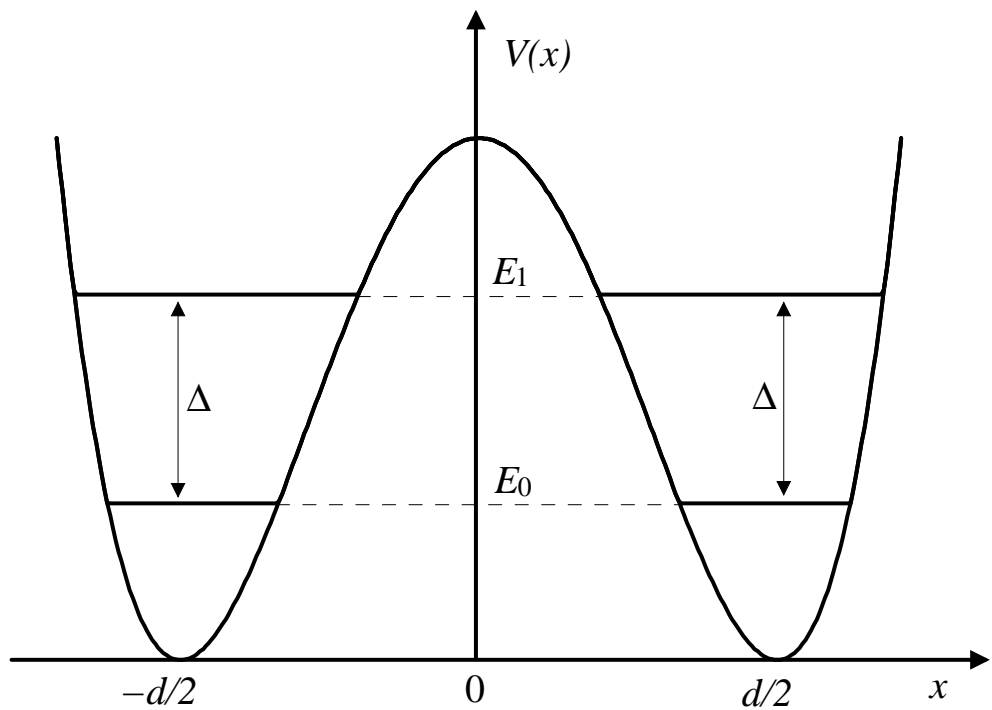
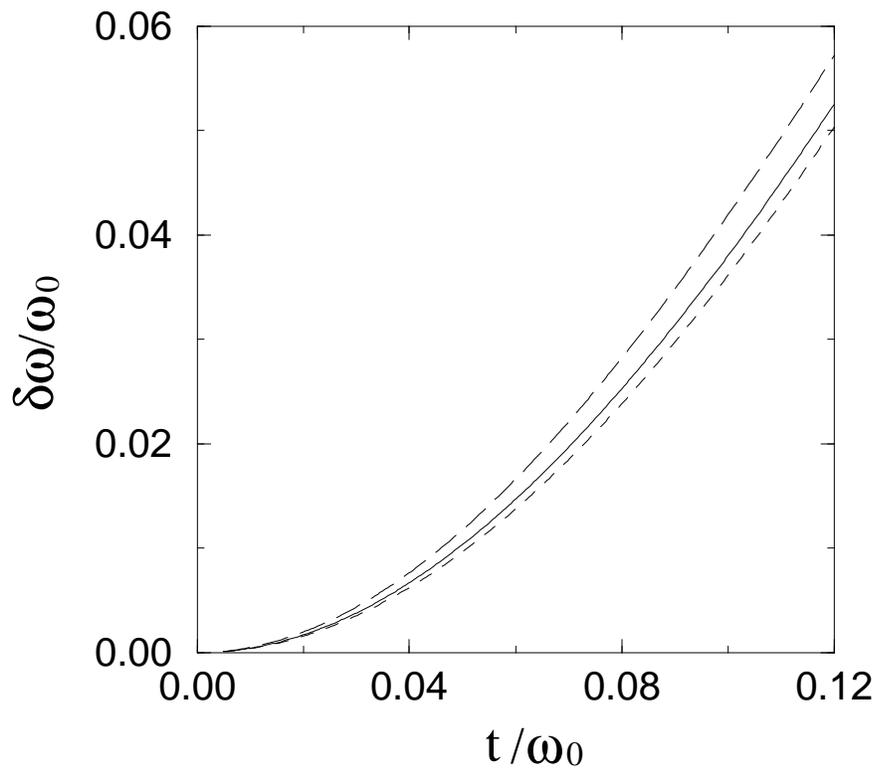
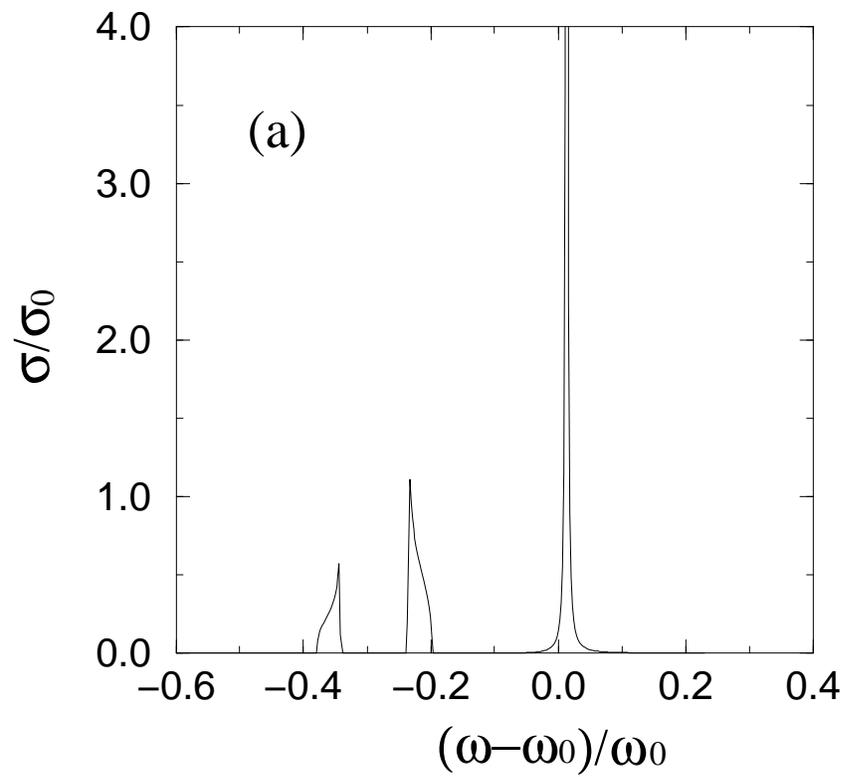
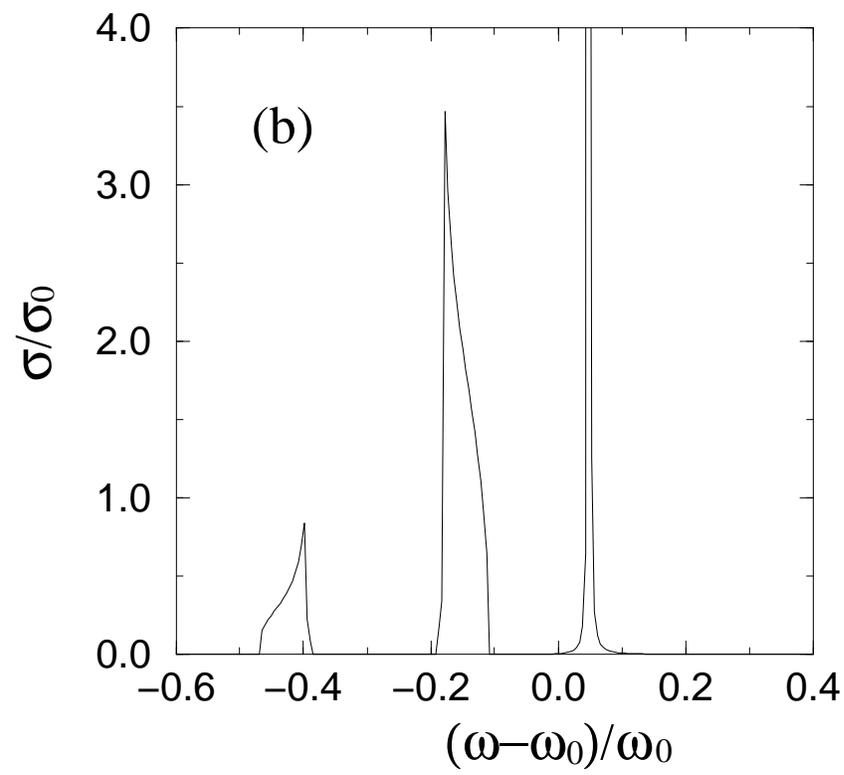
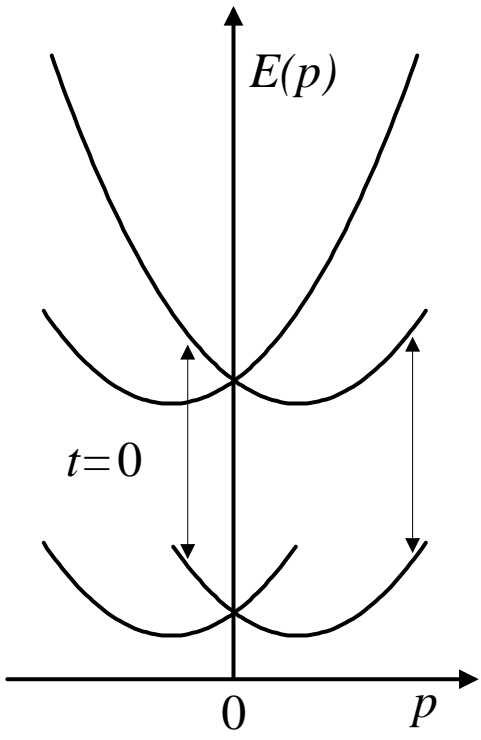
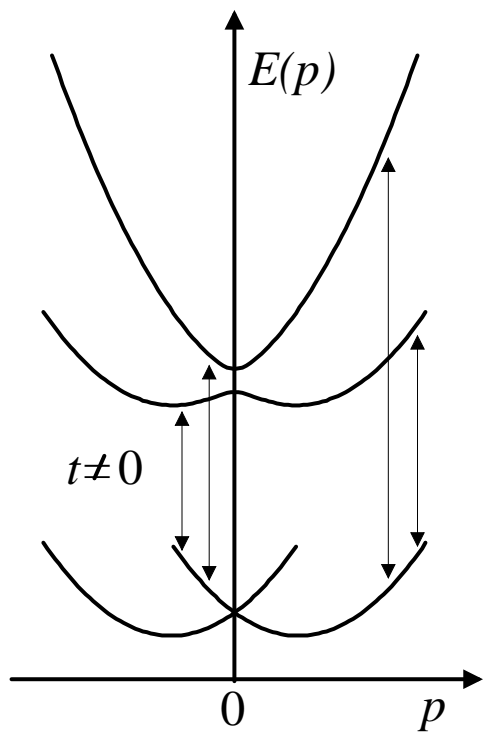
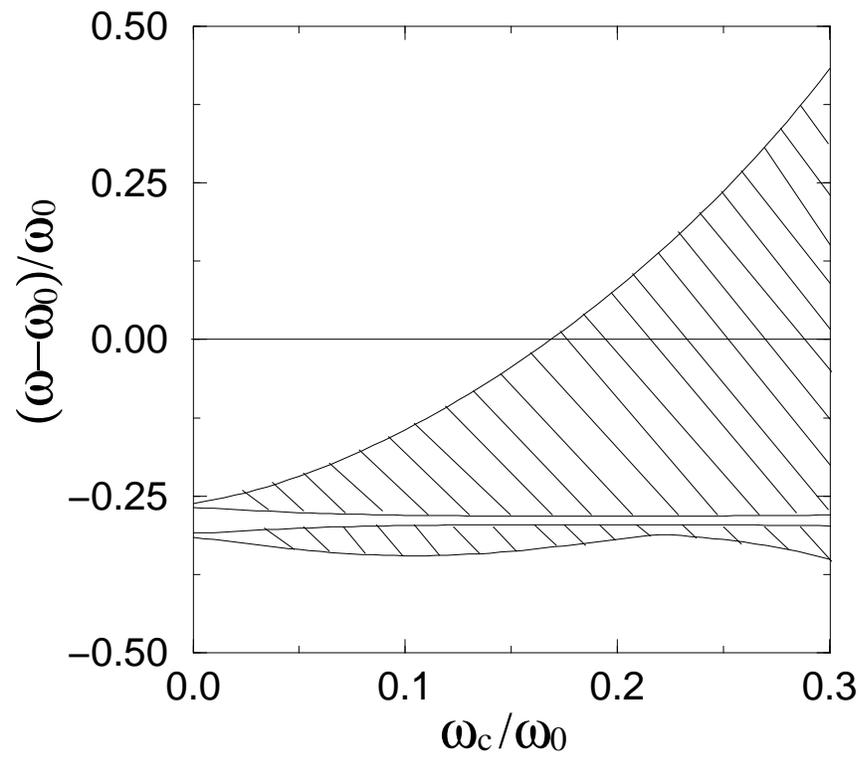
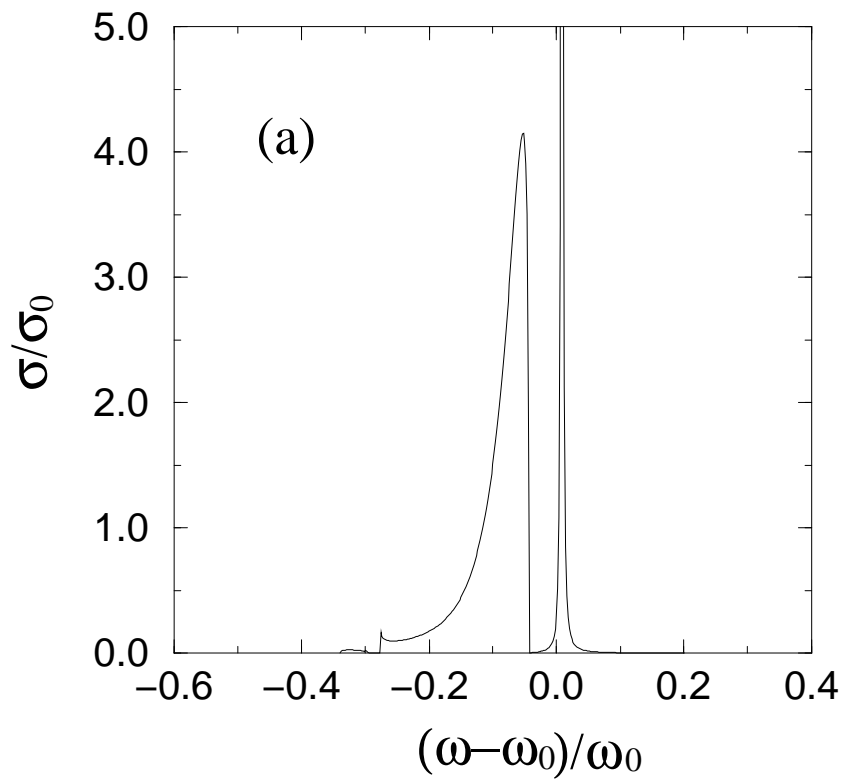
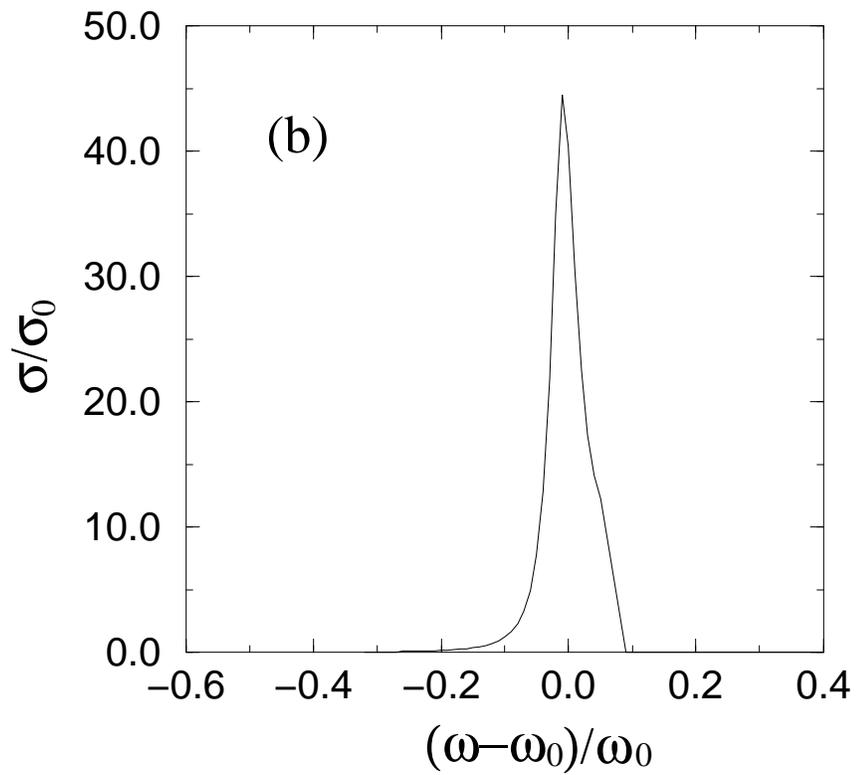
\end{document}